\documentclass[reprint,prb,superscriptaddress,showpacs]{revtex4-2}%
\usepackage{placeins}
\usepackage{xr-hyper}
\setcitestyle{maxcitenames=6}
\usepackage{gensymb}
\usepackage{graphicx}
\usepackage[autostyle]{csquotes}
\usepackage{xcolor}
\usepackage{amsmath}
\usepackage{amsfonts}
\usepackage{amssymb}
\usepackage{framed}
\usepackage{float}
\usepackage[normalem]{ulem}
\usepackage{lineno}
\usepackage{lipsum}
\providecommand{\U}[1]{\protect\rule{.1in}{.1in}}
\makeatletter
\renewcommand*{\fnum@figure}{{\normalfont\bfseries \figurename~\thefigure}}
\renewcommand*{\@caption@fignum@sep}{\textbf{ : }}
\makeatother
\usepackage{hyperref}
\hypersetup{
    colorlinks=true,
    linkcolor=blue,
    filecolor=magenta,      
    urlcolor=cyan,
}

\makeatother

\def\Tb166{TbMn$_6$Sn$_6$}
\def\Tm166{TmMn$_6$Sn$_6$}
\def\TmGa{TmMn$_6$Sn$_{4.2}$Ga$_{1.8}$}
\def\R166{RMn$_6$Sn$_6$}

\begin{document}
\title{Skyrmion Bubbles by Design in a Centrosymmetric Kagome Magnet}

\author{Mohamed El Gazzah}
\thanks{Equal contribution}
\email{melgazza@nd.edu} 
\affiliation{Department of Physics and Astronomy, University of Notre Dame, Notre Dame, IN 46556, USA}
\affiliation{Stavropoulos Center For Complex Quantum Matter, University of Notre Dame, Notre Dame, IN 46556, USA}

\author{Fehmi Sami Yasin}
\thanks{Equal contribution}
\thanks{Corresponding Author}
\email{yasinfs@ornl.gov}
\affiliation{Center for Nanophase Materials Sciences, Oak Ridge National Laboratory, Oak Ridge, TN 37830, USA}
\affiliation{RIKEN Center for Emergent Matter Science, Wako 351-0198, Japan}

\author{Sk Jamaluddin}
\affiliation{Department of Physics and Astronomy, University of Notre Dame, Notre Dame, IN 46556, USA}
\affiliation{Stavropoulos Center For Complex Quantum Matter, University of Notre Dame, Notre Dame, IN 46556, USA}

\author{Hari Bhandari}
\affiliation{Department of Physics and Astronomy, University of Notre Dame, Notre Dame, IN 46556, USA}
\affiliation{Stavropoulos Center For Complex Quantum Matter, University of Notre Dame, Notre Dame, IN 46556, USA}
\affiliation{Department of Physics and Astronomy, Rice University, Houston, Texas 77005, USA}

\author{Resham Babu Regmi}
\affiliation{Department of Physics and Astronomy, University of Notre Dame, Notre Dame, IN 46556, USA}
\affiliation{Stavropoulos Center For Complex Quantum Matter, University of Notre Dame, Notre Dame, IN 46556, USA}

\author{Xiuzhen Yu}
\affiliation{RIKEN Center for Emergent Matter Science, Wako 351-0198, Japan}
\affiliation{The Institute of Science Tokyo, Tokyo 152-8550, Japan}

\author{Nirmal J. Ghimire}
\thanks{Corresponding Author}
\email{nghimire@nd.edu}
\thanks{\\Notice: This manuscript has been authored by UT-Battelle, LLC, under contract DE-AC05-00OR22725 with the US Department of Energy (DOE). The US government retains and the publisher, by accepting the article for publication, acknowledges that the US government retains a nonexclusive, paid-up, irrevocable, worldwide license to publish or reproduce the published form of this manuscript, or allow others to do so, for US government purposes. DOE will provide public access to these results of federally sponsored research in accordance with the DOE Public Access Plan (https://www.energy.gov/doe-public-access-plan).}
\affiliation{Department of Physics and Astronomy, University of Notre Dame, Notre Dame, IN 46556, USA}
\affiliation{Stavropoulos Center For Complex Quantum Matter, University of Notre Dame, Notre Dame, IN 46556, USA}

\date{\today}

\begin{abstract}

\textbf{Topologically protected nanoscale spin textures, such as magnetic skyrmions, have attracted significant interest for spintronics applications. While skyrmions in noncentrosymmetric materials are known to be stabilized by Dzyaloshinskii-Moriya interaction (DMI), their deliberate design in centrosymmetric materials remains a challenge. This difficulty largely stems from the complexity of controlling magnetocrystalline anisotropy—a critical factor in the absence of DMI. Here, we demonstrate the chemical tuning of magnetocrystalline anisotropy in the centrosymmetric Kagome magnet TmMn$_\textbf{6}$Sn$_\textbf{6}$. The resulting compound exhibits a spin reorientation transition accompanied by an emergent skyrmion bubble lattice, confirmed by Lorentz transmission electron microscopy. Our findings overcome a key materials design challenge and open possibilities for deliberate design of skyrmionic textures in centrosymmetric systems.} 

\end{abstract} 

\maketitle




The identification, understanding, and manipulation of novel magnetic textures is essential for the characterization of new quantum materials which may compose future spin-based electronic devices \cite{fert2017magnetic,hirohata2020review,petrovic2021skyrmion}. Among these textures, magnetic skyrmions, vortex-like spin textures with integer topological charge $S$, have garnered particular attention due to their intriguing physical properties and technological potential \cite{mühlbauer2009skyrmion,yu2010real,nagaosa2013topological,finocchio2024roadmap}. Skyrmions were first discovered in 2009 in the noncentrosymmetric helimagnet MnSi \cite{mühlbauer2009skyrmion}, where their stabilization was governed by the antisymmetric Dzyaloshinskii-Moriya interaction (DMI), a consequence of broken inversion symmetry in the crystal lattice \cite{hayami2022skyrmion,zhou2025topological}. In such systems, a delicate balance between DMI, symmetric exchange interactions, and magnetocrystalline anisotropy facilitates the formation of these topologically protected spin textures. The excitement surrounding skyrmions intensified in 2010 when it was shown that they could be manipulated using electric current densities five orders of magnitude smaller than those required for conventional ferromagnetic metals and semiconductors \cite{jonietz2010spin} - highlighting their promise for low-power spintronic applications \cite{yu2012skyrmion,schulz2012emergent,lin2013particle,zhou2019magnetic}. 

\begin{figure*}[!ht]
    \begin{center}
        \includegraphics[scale = 0.72]{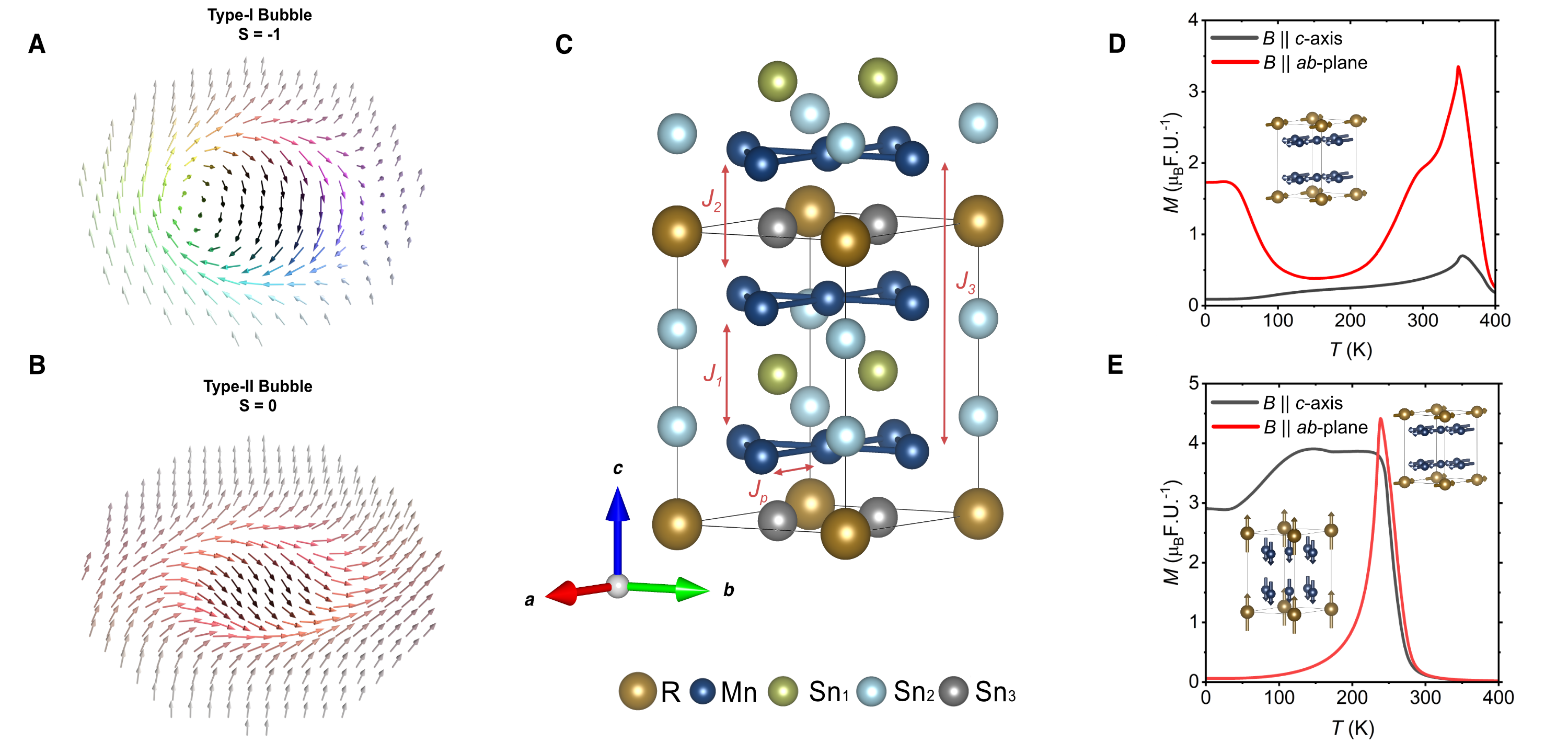}
        \caption{\textbf{Structural, magnetic, and topological characteristics of TmMn$_{\textbf{6}}$Sn$_{\textbf{6}}$}. (\textbf{A}-\textbf{B}) Schematics of (\textbf{A}) a Bloch Type-I magnetic skyrmion bubble and (\textbf{B}) a Type-II non-topological magnetic bubble. (\textbf{C}) Hexagonal P6/mmm (191) crystal structure of TmMn$_\textbf{6}$Sn$_\textbf{6}$ where the exchange interactions between the Kagome Mn layers are denoted $J_1$, $J_2$, and $J_3$, and the exchange interaction between Mn moments within the layer is denoted $J_{\text{P}}$. (\textbf{D}) Magnetic moment of the parent compound TmMn$_6$Sn$_6$ under an applied field of 0.1 T along both the $ab$-plane (red curve) and $c$-axis (black curve) showing a clear in-plane anisotropy throughout the full temperature range. (\textbf{E}) Magnetic moment of Ga-doped TmMn$_6$Sn$_6$, TmMn$_6$Sn$_{4.2}$Ga$_{1.8}$, under the same conditions with a clear spin reorientation appearing around 243 K.}
        \label{Fig1}
    \end{center}
\end{figure*}

Since then, skyrmions have been identified in a wide range of bulk materials and artificially layered heterostructures \cite{ezawa2010giant,lee2016synthesizing,woo2018current}. In the latter, interfacial DMI emerges from symmetry breaking at material interfaces, mirroring the stabilization mechanism seen in noncentrosymmetric helimagnets. Consequently, early research efforts focused mainly on DMI-driven skyrmion formation \cite{sampaio2013nucleation,jiang2015blowing,soumyanarayanan2017tunable}, under the assumption that DMI was a necessary ingredient. This assumption has been fundamentally challenged by recent discoveries of skyrmions in centrosymmetric materials systems that lack DMI entirely \cite{hayami2016bubble,hayami2021square,paddison2022magnetic,zuo2023spontaneous,chakrabartty2022tunable}. In these cases, skyrmion formation arises from a more complex interplay of magnetic and geometric frustration, higher-order exchange interactions, dipolar coupling, and magnetocrystalline anisotropy. These findings have expanded the landscape of materials that can host skyrmions and raised important questions about the deeper mechanisms that govern their stabilization \cite{tokura2020magnetic}.

Despite these advances, a unifying characteristic across all known skyrmion-hosting systems is that their emergence has been passive: they have been stabilized solely by the already present interactions within the material. This leads to a pivotal question: can we move beyond discovery and toward the active engineering of skyrmions on demand? Achieving such control would represent a remarkable step in topological magnetism, enabling the deliberate design and precise manipulation of skyrmionic textures for future spintronics applications. This would not only allow for control over the size and density of skyrmions considered important for technological applications \cite{kang2016skyrmion} but also open avenues for coupling them with electronic topological states, potentially leading to more exotic phenomena \cite{martin2012majorana,freimuth2013phase,lado2015quantum,petrovic2021skyrmion,paul2021topological,petrovic2024colloquium}. These possibilities echo the theoretical predictions of real- and momentum-space topology interplay, such as those proposed in the $A$-phase of superfluid helium-3 \cite{volovik2003universe,volovik2001interplay} and more recently discussed theoretically in condensed matter systems \cite{martin2012majorana,nagaosa2012gauge,freimuth2013phase,paul2021topological}. However, a major challenge in the proposed strategies for engineering skyrmions in centrosymmetric systems \cite{wang2020skyrmion} lies in the control of magnetocrystalline anisotropy \cite{karube2022doping}. As a result, previous efforts have primarily focused on materials incorporating weakly anisotropic rare-earth elements such as Gd \cite{kurumaji2019skyrmion,khanh2020nanometric,hirschberger2019skyrmion} and Eu \cite{kakihana2019unique,moya2022incommensurate,zhang2021giant,gen2023rhombic}.

In this article, we demonstrate a controlled design of skyrmion bubbles and their manipulation from Type-I (integer topological charge, $S=\frac{1}{4\pi}\int\int\textbf{m} \cdot \left(\frac{\partial \mathbf{m}}{\partial x} \times \frac{\partial \mathbf{m}}{\partial y} \right) dx\ dy = -1$, where $\mathbf{m}=\frac{\mathbf{M}}{|\mathbf{M}|}$ is the normalized magnetization) to Type-II (non-topological, $S = 0$) shown in Figs. \ref{Fig1}(A-B)—in the centrosymmetric kagome lattice compound \Tm166 primarily by engineering the magnetocrystalline anisotropy through a targeted chemical substitution. We use real-space imaging via Lorentz transmission electron microscopy to show that the stabilization of a skyrmion bubble lattice occurs immediately below the engineered spin reorientation transition, where competing anisotropies from the rare-earth and transition-metal sublattices are maximal. Our results reveal a dense, tunable skyrmionic bubble lattice phase that persists across a wide temperature and field window. Furthermore, we observe spontaneous skyrmion bubble helicity switching and the coexistence of opposite helicities- features that highlight the complex and highly sensitive energy landscape of these spin textures as well as the apparent degeneracy of both bubble helicities. Our findings offer a new pathway: by harnessing competing anisotropies through chemical substitution, one can tune the system to the optimal conditions of realizing a spin reorientation transition. Crucially, the magnetic anisotropy can be precisely tuned via chemical substitution, enabling the deliberate induction of a spin reorientation transition resulting in the stabilization of topological spin textures in centrosymmetric systems by design. This approach not only offers refined control over skyrmion topology and stability but also provides a versatile platform for integrating skyrmions phases with broader quantum phenomena.

\subsection*{Spin reorientation within \Tm166}

\Tm166 belongs to the broader family of hexagonal \R166 compounds, which have recently garnered significant attention for their complex and tunable magnetic behavior. This richness arises from a delicate interplay of competing interlayer exchange interactions—$J_1$, $J_2$, and $J_3$—as illustrated in Fig. \ref{Fig1}(C) \cite{ghimire2020competing,jones2024origin,siegfried2022magnetization}. Within this family, YMn$_6$Sn$_6$ has been proposed to host a thermally stabilized dynamical skyrmion phase, believed to underlie its observed topological Hall response \cite{ghimire2020competing}. Similarly, in \Tb166, a biskyrmion lattice has been reported to emerge exclusively near a spontaneous spin reorientation transition \cite{li2023discovery}, suggesting that such a reorientation may be key to stabilizing topological spin textures.

Recent work has shown that the spin reorientation in \Tb166 originates from competition between the magnetic anisotropies of the Mn and Tb sublattices, modulated by thermally driven spin fluctuations consistent with Mermin-Wagner physics \cite{jones2024origin}. Both sublattices order magnetically at about 420 K, but at elevated temperatures, the stronger in-plane anisotropy of Mn sub-lattice dominates. Below 310 K, however, the out-of-plane anisotropy of the Tb moments prevails, leading to a spontaneous spin reorientation of both Mn and Tb spins along the $c$-axis. In contrast, \Tm166 exhibits robust in-plane anisotropy over the entire temperature range \cite{venturini1996incommensurate,canepa2005magnetisation}, as evidenced in the temperature-dependent magnetization data (Fig. \ref{Fig1}(D)). Remarkably, substitution of Ga into \Tm166 induces a clear spin reorientation below 243 K (Fig. \ref{Fig1}(E)). To probe the resulting magnetic textures, we employed high-resolution imaging of magnetic domains in Ga-doped \Tm166 using Lorentz Transmission Electron Microscopy (LTEM) and Differential Phase Contrast Scanning Transmission Electron Microscopy (DPC-STEM).

\begin{figure*}[!ht]
\begin{center}
\includegraphics[scale=0.13]{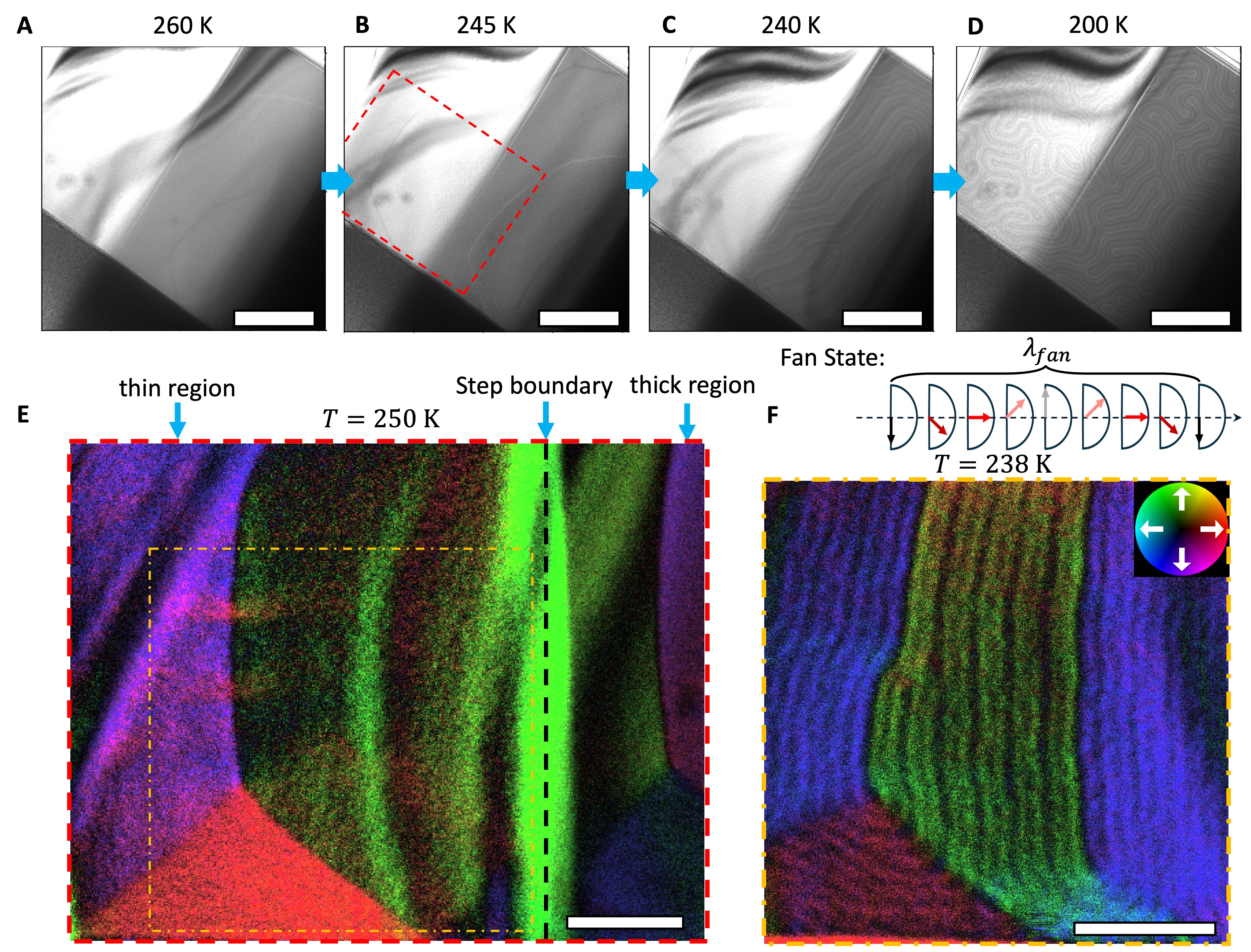}
    \caption{\textbf{Real-space evolution of magnetic textures across the spin reorientation transition.} (\textbf{A-D}): Selected Lorentz transmission electron micrographs (LTEM) of magnetic textures in TmMn$_6$Sn$_{4.2}$Ga$_{1.8}$ during zero-field cooling from (\textbf{A}) 260 K to (\textbf{B}) 245 K, (\textbf{C}) 240 K, and (\textbf{D}) 200 K. Scalebars, $1\ \mu\textrm{m}$.(\textbf{E}) Differential Phase Contrast Scanning Transmission Electron Microscopy (DPC-STEM) image of the selected region indicated by the dashed red rectangle outlined in (B) taken at 250 K and zero field, revealing ferromagnetic in-plane domains. The black dashed line indicated the step-like increase in thickness separating the thin region on the left-hand-side from the thick region on the right-hand-side. (\textbf{G}) DPC-STEM image of the selected area in the thin region indicated by the yellow dot-dashed rectangle outlined in (E) at the spin reorientation temperature 238 K, revealing a magnetic fan state (shown schematically at the top of panel F where the magnetization periodically twists from out-of-plane to in-plane to out-of-plane before reversing direction and sweeping a `fan' shape over the period $\lambda_{\text{fan}}$) which acts as the intermediary state between in- and out-of-plane spin states. Scalebars, $0.4\ \mu\textrm{m}$.} \label{Fig2}
    \end{center}
\end{figure*}

LTEM micrographs of a $c$-axis wedge plate of TmMn$_6$Sn$_{4.2}$Ga$_{1.8}$, with thicknesses of 91 nm in the darker region and 43 nm in the lighter region, are shown across temperatures from 260 K to 200 K in Figs. \ref{Fig2}(A-D). At 260 K, the thicker region reveals well-defined in-plane domains, identified through the existence of a dark-intensity $180 \degree$ domain wall near the center of the thin plate which branches to form two $90\degree$ domain walls near the bottom edge of the LTEM micrograph. On the other hand, the thinner region shows no visible magnetic domain walls, indicating the presence of either a single magnetic domain or a paramagnetic state. Above the spin reorientation temperature $T_{\text{SR}}$ (at 245 K), in-plane domains appear prominently in both thick and thin regions, confirming strong in-plane anisotropy. At 240 K, the thicker region transitions to predominantly striped domains, indicative of a shift to out-of-plane anisotropy, while the thinner region retains in-plane magnetic domains. Further cooling to 200 K, results in both regions transitioning from in-plane to out-of-plane anisotropy along the $c$-axis, producing complex maze-like striped domains across the entire lamella shown in Fig. \ref{Fig2}(D).

To investigate the spin structure around the spin reorientation temperature, we performed DPC-STEM. The in-plane magnetic induction maps at $T=250\ \textrm{K}$ and $T=238\ \textrm{K}$ are shown in Figs. \ref{Fig2}(E-F). At 250 K, we find in-plane magnetic domains in both regions of the lamella separated by either $180 \degree$ (e.g., between purple and green domains) or $90 \degree$ (e.g., between purple/green and red/blue domains) domain walls. At 238 K in the thin region indicated by the yellow dot-dashed box in Fig. \ref{Fig2}(E), a periodic structure emerges that DPC-STEM reveals in Fig. \ref{Fig2}({F}) to be a fan state. The spins in this fan state trace a fan-like shape across the period ($\lambda_{\text{fan}}$) of the texture, as shown in the schematic above Fig. \ref{Fig2}({F}). These magnetic fan states have modulation $\textbf{q}$-vectors that align perpendicular to either the edge (for the red/blue domains) or the $180 \degree$ domain walls. As the temperature decreases further and the uniaxial anisotropy strengthens, the fan-like spin modulations reorient smoothly into the labyrinth domains separated by $180 \degree$ domain walls shown in Fig. \ref{Fig2}({D}).

\hspace*{-0.8cm}\begin{figure*}[!ht]
\begin{center}
\includegraphics[scale=0.12]{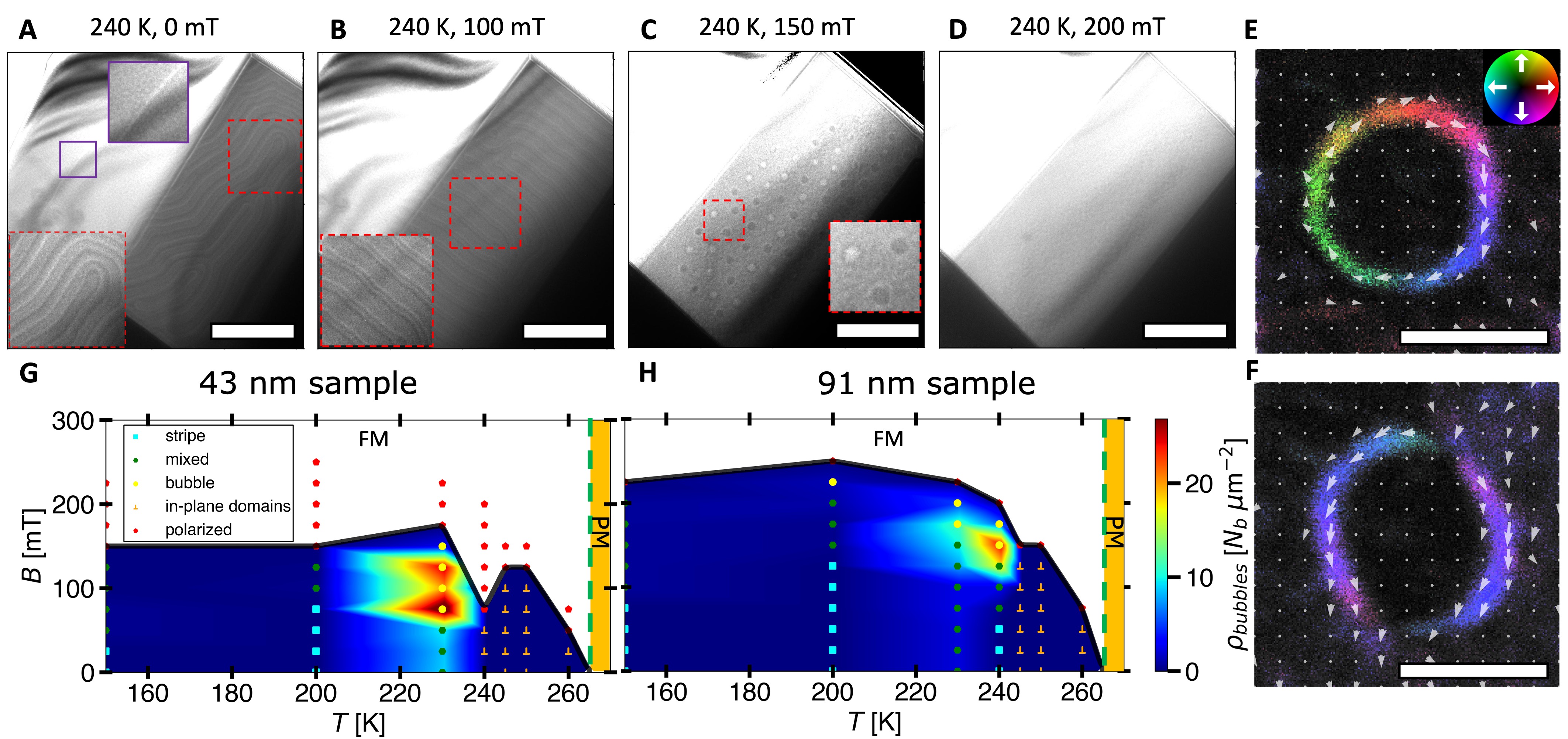}
    \caption{\textbf{Thickness dependence on the magnetic field - temperature magnetic state diagrams of TmMn$_\textbf{6}$Sn$_{\textbf{4.2}}$Ga$_{\textbf{1.8}}$.} ({\textbf{A}-\textbf{D}}) Selected LTEM micrographs of a thin and thick region in the TmMn$_6$Sn$_{4.2}$Ga$_{1.8}$ lamella depicting the magnetic states at $T = 240$ K under external magnetic fields of (\textbf{A}) $B_{\text{ext}} = 0$ mT, (\textbf{B}) $B_{\text{ext}} = 100$ mT, (\textbf{C}) $B_{\text{ext}} = 150$ mT, and (\textbf{D}) $B_{\text{ext}} = 200$ mT. Scalebars, $1\ \mu\textrm{m}$. The magnetic states evolve from (\textbf{A}) out-of-plane domains with maze-like modulations and $180\degree$ domain walls at zero field to a (\textbf{B}) single q-vector stripe state at $B=100\ \textrm{mT}$, likely due to the presence of a slight sample tilt and therefore a non-zero in-plane component of the magnetic field relative to the [001] crystal axis. Skyrmionic bubbles emerge at (\textbf{C}) $B=150\ \textrm{mT}$ and the spins polarize into a ferromagnetic state above $B=200\ \textrm{mT}$. (\textbf{E} - \textbf{F}) In-plane magnetic induction maps of a \textbf{E} Type-I skyrmion bubble and a \textbf{F} Type-II non-topological bubble measured using DPC-STEM. Scalebars, $0.1\ \mu\textrm{m}$.(\textbf{G-H}) The temperature $T\ [\textrm{K}]$ versus applied magnetic field $B\ [\textrm{mT}]$ magnetic state diagrams of the (\textbf{G}) thin $43\ \textrm{nm}$ and (\textbf{H}) thick $91\ \textrm{nm}$ regions of the thin plate. The magnetic bubble density $\rho_{\text{bubbles}}$ is plotted in color, and the uniaxial anisotropic stripe domains, mixed stripe and magnetic bubbles, magnetic bubble lattice, in-plane domains, and polarized states are indicated by cyan squares, green hexagons, yellow circles, orange `$\perp$,' and red pentagons, respectively. The curie temperature is indicated by a green dashed vertical line and the ferromagnetic (polarized) and paramagnetic states are labeled as `FM' and `PM' respectively.} \label{Fig3}
    \end{center}
\end{figure*}

\subsection*{Emergence of skyrmionic bubbles}
Below the spin reorientation temperature, the magnetic textures resemble those of prototypical magnets with uniaxial anisotropy. As such, the application of an external magnetic field parallel to the $c$-axis drives the formation of type-I and type-II bubbles, illustrated in the magnetic field sweep in Figs. \ref{Fig3}({A-D}) at 240 K. At zero field (Fig. \ref{Fig3}(A)), the thicker region exhibits maze-like magnetic domains with narrow domain walls. Upon application of a $B=100\ \textrm{mT}$ magnetic field, the stripe domains widen in the regions with magnetization aligned parallel to the applied field and shrink in the anti-parallel domains as shown in Fig. \ref{Fig3}({B}). At $B=150\ \textrm{mT}$, skyrmion bubbles stabilize and form a loose triangular lattice (Fig. \ref{Fig3}{(C)}. At $B=200\ \textrm{mT}$, the lamella saturates with spins polarized along the applied magnetic field. Similar behavior is seen in the thinner region (Fig. \ref{fig:FigS2}). In-plane magnetic induction maps of type-I and type-II Bloch skyrmion bubbles are shown in Figs. \ref{Fig3}({E}) and \ref{Fig3}({F}), respectively. These two bubble types are observed in both the thick and thin regions, and so is their reversible transformation (between type-I and type-II bubbles) via the application of an oblique magnetic field (sample tilt), similar to prior work \cite{yu2012magnetic}.

The magnetic state diagrams for the $d=43\ \textrm{nm}$ and $d=91\ \textrm{nm}$ thick regions are shown in Figs. \ref{Fig3}({G}) and \ref{Fig3}({H}). The spins order from a paramagnetic state labelled `PM' in the yellow shaded regions on the right-hand-side of the green dashed line indicating the Curie temperature, to in-plane domains (indicated by orange $\perp$ markers), which become polarized (marked by red pentagons) under sufficiently high applied magnetic fields. This behavior is consistent in both the thick and thin regions, although a higher magnetic field is required to polarize the larger volume of spins in the thicker sample. At the spin reorientation temperature and zero field, a fan state was observed as discussed in Fig. \ref{Fig2}. Below this temperature, we observe a range of magnetic structures, including striped domains, bubbles, and mixed states consisting of bubbles and striped domains marked in Figs. \ref{Fig3}(G-H) as cyan squares, yellow circles, and green hexagons. Striped domains emerge at lower magnetic fields and transition into a mixed state with increasing field strength before fully transforming into bubbles. Further increases in the magnetic field lead to saturation of the thin plate into a ferromagnetic state labelled as `FM' in Figs. \ref{Fig3}(G-H).

In the 43-nm region, a high density of magnetic bubbles, marked by yellow dots, is observed at 230 K at fields ranging from 60 mT to 140 mT. In contrast, the 91-nm region exhibits a lower bubble density, forming at higher fields over a broader temperature range: at 240 K with 180 mT, at 230 K with 180-200 mT, and at 200 K with 220 mT. This behavior qualitatively follows what we would expect from Kittel's law \cite{kittel1946theory}, which states that the characteristic domain width $D$ scales with the square root of the thickness. To account for this, we rescaled the magnetic bubble density in the thicker region (Fig. \ref{Fig3}(H)) by $D(d=43\ \textrm{nm}) / D(d=91\ \textrm{nm}) \propto 91\ \textrm{nm} / 43\ \textrm{nm} = 2.12$, so that the two $\rho_{\text{bubbles}}$ may be compared directly \cite{kittel1946theory}.

The thickness-dependent behavior highlights the critical interplay between the dipolar interactions and anisotropy in stabilizing topological magnetic bubbles \cite{heigl2021dipolar}. In thinner samples, the reduced volume enhances dipolar interactions due to decreased screening effects, promoting the formation and stability of magnetic bubbles. However, uniaxial anisotropy plays an equally crucial role, as it favors out-of-plane spin alignment, which is necessary for stabilizing these bubbles. To this extent, a quality factor Q, can be defined as the ratio of anisotropy energy, $K$, and the dipolar interaction, $2\pi M^2$, $Q = \frac{K}{2\pi M^2}$. Bubble formation is only possible for $Q>1$, indicating that uniaxial anisotropy must dominate over dipolar interactions. This relationship is particularly evident in systems with reduced thickness, where the enhanced dipolar interactions, combined with sufficient uniaxial anisotropy, allow for the emergence of topological textures \cite{hassan2024dipolar,heigl2021dipolar,jefremovas2024role}.

As discussed, the magnetic induction maps reveal two distinct magnetic bubbles stabilized in TmMn$_6$Sn$_{4.2}$Ga$_{1.8}$, each with unique topological and magnetic characteristics. Type-I bubbles (depicted in Fig. \ref{Fig1}(A) and shown in Fig. \ref{Fig3}({E})), are characterized by a circular domain with a dark core surrounded by a perimeter where the spins undergo a full $360 \degree$ rotation. The spin configuration in Type-I bubbles wraps the unit sphere once, making them topologically non-trivial and akin to skyrmions, imparting stability against thermal fluctuations and external perturbations. Such topological protection arises because the spins cannot be continuously deformed to a trivial state without breaking the continuity of the magnetization, making Type-I bubbles particularly robust in varied magnetic environments. Typically, the topological Hall effect (THE) arises from the real-space Berry curvature induced by the emergent magnetic field of nontrivial spin texture such as skyrmions. However, the present case deviates from that expectation. We suspect this is due to the relatively large size of the skyrmion bubbles ($\approx 100\ \textrm{nm}$), which reduces their emergent magnetic field that contributes to the topological Hall effect. Furthermore, the skyrmion bubble lattice is observed only in very thin samples ($d\approx 50\ \textrm{nm}$), which poses significant challenges for reliably measuring the Hall effect in such ultra thin specimens.

On the other hand, Type-II bubbles (illustrated in Fig. \ref{Fig1}(B) and imaged in Fig. \ref{Fig3}(F)) exhibit a non-topological structure. Here, the spins do not complete a full rotation around the perimeter; instead, they align symmetrically on opposite sides, converging at the Bloch lines, which are Néel-like interfaces separating the two Bloch-type domain wall segments. This configuration results in a state topologically equivalent to a striped domain or a polarized domain. Interestingly, we find that Type-II bubbles are stabilized under the application of small in-plane magnetic fields, essential for balancing the competing magnetic anisotropies and dipolar interactions. This necessity for an oblique field confirms that Type-II bubbles are more sensitive to external field orientation and magnitude, unlike the more robust Type-I bubbles.

The distinct magnetic properties of Type-I and Type-II bubbles have significant implications. The topological stability of Type-I bubbles makes them promising candidates for information storage and manipulation in spintronics, where resilience to disturbances is crucial. In contrast, Type-II bubbles, with their sensitivity to in-plane fields and anisotropies, could be utilized in applications requiring rapid switching or modulation of magnetic states \cite{vzutic2004spintronics,zhou2025topological,luo2021skyrmion}. The ability to stabilize both through careful control of doping and external magnetic fields within the TmMn$_6$Sn$_{4.2}$Ga$_{1.8}$ system showcases the versatility of the RMn$_6$Sn$_6$ family. This tunable magnetic behavior provides a powerful framework for engineering diverse magnetic textures and exploring new functional properties in quantum materials.

\subsection*{Thermally driven dynamic bubble helicity flipping}

\begin{figure}[!ht]
\begin{center}
\includegraphics[scale=0.19]{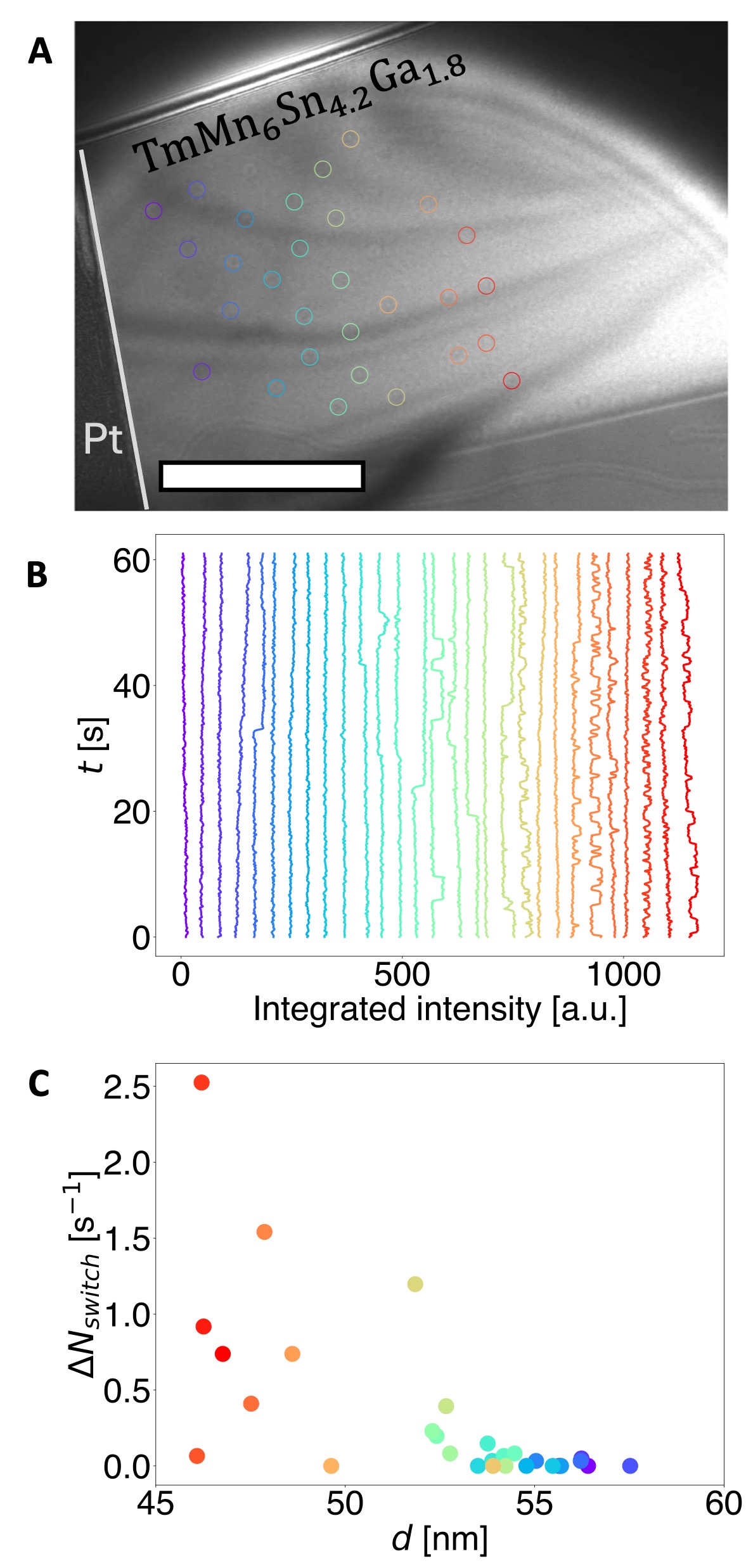}
    \caption{\textbf{Thickness-dependent helicity switching of magnetic bubbles.} (\textbf{A-C}): Helicity switching of magnetic bubbles in TmMn$_6$Sn$_{4.2}$Ga$_{1.8}$ at $B_{\text{ext}} = 125$ mT and $T = 210$ K. (\textbf{A}) LTEM micrograph showing the thermodynamically stable magnetic bubble lattice state, with bubbles selected for helicity switching analysis marked by solid line circles in different colors. Scalebar is $1\ \mathrm{\mu m}$. (\textbf{B}) Time evolution of the integrated intensity within the selected bubbles, where the step-like changes in intensity indicate helicity switching events. (\textbf{C}) The rate of helicity switching $\Delta N_{\text{switch}}$ as a function of thickness averaged over the bubbles' in-plane area, illustrating the impact of lamella thickness on helicity switching rates.}
    \label{Fig4}
    \end{center}
\end{figure}

Interestingly, during the real space microscopy experiments, we found clear signatures of helicity switching in Type-I magnetic bubbles. Given the inherent topological protection of these bubbles, the helicity switching under applied field during LTEM imaging is particularly intriguing. While the topological nature of Type-I bubbles makes them robust against collapse into a topologically trivial state, their helicity, or the chirality of the domain wall, is susceptible to dynamic switching between left-handed and right-handed configurations, while preserving integer topological charge, as both configurations are energetically equivalent \cite{kong2024diverse}. 

To quantify this phenomenon in TmMn$_6$Sn$_{4.2}$Ga$_{1.8}$, we tracked the helicity switching rate of several bubbles across the thin region of the FIB-fabricated lamella which has a wedge-like thickness dependence. A defocussed LTEM micrograph is shown in Fig. \ref{Fig4}(A) with circular markers of varying colors (thickness increases from cooler to warmer colors) drawn around the skyrmion bubbles selected for analysis. The thickness map of the lamella, measured using electron energy loss spectroscopy (EELS), is presented in Fig. \ref{fig:FigS2} \cite{egerton1987measurement}. Fig. \ref{Fig4}(B) shows step-like changes in the integrated intensity within the bubbles central core region, representing discrete helicity switching events because the vortex-like in-plane magnetic field focuses the electrons radially outward or inward depending on helicity. The rate of helicity switching $\Delta N_{\text{switch}}$ is plotted as a function of average sample thickness at the bubbles' locations in Fig. \ref{Fig4}(C). The strong thickness dependence suggests that the energy required to switch the bubble helicity—likely through the formation of a Bloch line—decreases as the thickness is reduced.

One possible explanation for this behavior is the increased dipolar energy in thinner regions, where reduced screening enhances the influence of dipolar fields on the local magnetic environment. This increased dipolar interaction reduces the energy barrier for certain magnetic events, leading to an increase in thermally-activated helicity switching events. This observation aligns with our findings that helicity switches are more frequent in thinner regions of the sample, indicating a correlation between reduced volume, increased magnetic flexibility (propensity of spin textures to respond to perturbations), and helicity dynamics. The maximum switching rate reaches 2.6 switches per second in the thinner region.

These observations are similar to the magnetic bubble helicity switching observed by Yu et al. \cite{yu2016thermally} in a [001] thin plate of hexaferrite crystal, BaFe$_{12-x-0.005}$Sc$_x$Mg$_{0.005}$O$_{19}$ (BFSMO). In their study, helicity reversals in skyrmion bubbles were thermally activated by globally heating the sample near its Curie temperature, $T_{\text{C}} \approx 450$ $K$. At room temperature, the skyrmion bubbles coexist with the helical phase and persist to high temperatures, allowing for the study of helicity dynamics under thermal perturbation. Yu et al. reported that the helicity-switching speed exceeded the frame rate (30 fps) of their LTEM movie, corresponding to at least 30 switches per second—significantly faster than our observation. This highlights the significant influence of thermal perturbations on helicity switching, suggesting the helicity dynamics we observe may result from weaker, more localized heating effects caused by the electron beam during LTEM imaging.
\subsection*{Conclusions}
By precisely manipulating magnetic anisotropy via Ga doping in TmMn$_6$Sn$_6$, we induce a notable spin reorientation in TmMn$_6$Sn$_{4.2}$Ga$_{1.8}$, leading to the stabilization of both topological (Type-I) and non-topological (Type-II) magnetic bubbles. The helicity reversal in skyrmion bubbles illuminates the role of thickness in modulating the necessary interactions. This capability of leveraging chemical doping to fine-tune anisotropy and generate a spin reorientation, transforms our ability to design and engineer topological spin textures. As we continue to investigate similar systems, different dopants or concentrations may further refine the dimensions and dynamics of magnetic bubbles, or even reveal more exotic phenomena within the broader Kagome family and beyond.

\section*{Methods}
\noindent
\textbf{Crystal growth and structural characterization:} We grew single crystals of TmMn$_6$Sn$_{6-x}$Ga$_x$ through the self-flux method using Sn as a flux. Tm pieces (Alfa Aesar; 99.9\%), Mn pieces (Alfa Aesar; 99.95\%), Sn shots (Alfa Aesar; 99.999\%),and Ga pieces (Alfa Aesar; X\%) were added into a 2 mL aluminum oxide crucible in a molar ratio of 1: 6: 20 for \Tm166 and 1: 6: 15.5: 4.5 for \TmGa, and were sealed in a fused silica ampule under vacuum. The sealed ampule was heated to 1150 $^{\circ}$C over 10 hours, kept at 1150 $^{\circ}$C for 10 hours, and then cooled to 600 $^{\circ}$C at a rate of 5 $^{\circ}$C/h before the tube was centrifuged. Large, hexagonal, plate-like crystals (up to 50 mg) were obtained.

Although the initial nominal Ga concentration was $x = 4.5$, structural and compositional analyses revealed the actual Ga content to be $x = 1.8$, as determined through powder X-ray diffraction (XRD) and energy-dispersive X-ray spectroscopy (EDX). Crystals were cleaned of residual Sn flux prior to characterization. Phase purity and structural parameters were confirmed by Rietveld refinement of powder XRD patterns collected from crushed single crystals using a Rigaku Miniflex diffractometer, with refinement carried out using the FULLPROF suite. Single-crystal X-ray diffraction further validated the refined lattice parameters. Elemental composition and homogeneity were confirmed via EDX mapping using a JEOL JSM-IT500HRLV scanning electron microscope equipped with an Octane Elect Plus EDX detector. Measurements were performed at an accelerating voltage of 15 kV, with elemental ratios determined within 7\% error.\\

\noindent
\textbf{Magnetic measurements:} We measured temperature-dependent susceptibility with 0.1 T applied magnetic field and DC magnetization using the Quantum Design Dynacool Physical Property Measurement System (PPMS) with a 9-T magnet. Both measurements were done along the $ab$-plane and along the $c$-axis. The AC Measurement System (ACMS) option was used for both magnetization measurements.\\

\noindent
\textbf{Lamella preparation}
After confirming the desired composition, we prepared electron-transparent lamellae for LTEM analysis via Ga-ion focused ion beam (FIB) on a Thermo Fisher Nova 200 Dual Beam FIB. After thinning the entire lamella to a thickness of 91 nm, we continued thinning half of the lamella to $\approx43$ nm to study the effect of thickness on spin texture stabilization. The resulting thickness profile is shown in Fig. \ref{fig:FigS2}.\\

\noindent
\textbf{Lorentz transmission electron microscopy:} We performed the electron microscopy experiments in low-magnification transmission electron microscopy (TEM) mode with the objective and mini-condenser lenses turned off (Lorentz mode) in both a Talos F200X S/TEM under $200\ \textrm{keV}$ acceleration voltage and a JEOL NeoARM. For the magnetic state diagram mapping shown in Figs. \ref{Fig3}(G-H), we cooled the samples to $T=150 \ \textrm{K}$ from room temperature using a Gatan 636 double tilt liquid nitrogen cooling holder and a custom-built Gatan liquid nitrogen TEM holder with temperature control. Prior to applying a magnetic field $B_{\text{ext}}$ by exciting the objective lens current, we oriented the sample's $c$-axis to be parallel with the optical axis using selected area electron diffraction. We increased $B_{\text{ext}}$ until the magnetic contrast in the LTEM image disappeared, which we interpreted as the polarization of spins into a ferromagnetic state. We then increased the temperature to the next highest one indicated by the markers in Figs. \ref{Fig3}(G-H) and decreased $B_{\text{ext}}$ to zero before performing a field sweep to saturation at the target temperature. We repeated this process at every temperature indicated by markers in Figs. \ref{Fig3}(G-H). We used scanning transmission electron microscopy differential phase contrast (STEM-DPC) imaging in both the Talos F200X S/TEM and JEOL NeoARM equipped with an annular pixelated detector.\\

\noindent
\textbf{Thickness mapping via electron energy loss spectroscopy:} We performed electron energy loss spectroscopy on a JEOL NeoARM at 200 keV beam energy with a collection angle of $\beta = 48.2\ \textrm{mrad}$. We estimated the electron's mean free path length based on the Mean Atomic Number $Z = 37.29$ to be $\lambda = 71.64\ \textrm{nm}$, which we used to generate the thickness maps in Fig. \ref{fig:FigS2}.

\section*{Acknowledgments}
N.J.G. acknowledges the support from the NSF CAREER award DMR-2343536. X.Y. acknowledges Grants-In-Aid for Scientific Research (A) (Grant No. 19H00660) from the Japan Society for the Promotion of Science (JSPS) and the Japan Science and Technology Agency (JST) CREST program (Grant No.  JPMJCR20T1), Japan. Work conducted as part of a user project at the Center for Nanophase Materials Sciences, a U.S. Department of Energy Office of Science User Facility at Oak Ridge National Laboratory. Research sponsored by the Laboratory Directed Research and Development Program of Oak Ridge National Laboratory, managed by UT-Battelle, LLC, for the US Department of Energy.

\section*{Author contributions}
F.S.Y., N.J.G. and X.Y. conceived the idea. F.S.Y. and N.J.G. coordinated the project. M.E.G., and H.B. grew single crystals. M.E.G., and R.B.R. performed structural characterizations, and physical properties measurements. F.S.Y. carried out LTEM and DPC-STEM measurements. M.E.G and J.S. independently reproduced LTEM measurements. F.S.Y., M.E.G., and N.J.G.
wrote the manuscript. All authors contributed to the discussion
of the results.\\


%

\widetext
\begin{center}
\pagebreak
\hspace{0pt}
\vfill
\textbf{\large Supplementary Material}
\vfill
\hspace{0pt}
\end{center}
\FloatBarrier

\setcounter{equation}{0}
\setcounter{figure}{0}
\setcounter{table}{0}
\setcounter{page}{1}
\makeatletter
\renewcommand\thesection{S\arabic{section}}
\renewcommand{\theequation}{S\arabic{equation}}
\renewcommand{\thetable}{S\arabic{table}}
\renewcommand\thefigure{S\arabic{figure}}
\renewcommand{\theHtable}{S\thetable}
\renewcommand{\theHfigure}{S\thefigure}

\section{Crystal Structure and Powder X-Ray Diffraction Analysis}

\begin{figure}[H]
\begin{center}
\includegraphics[scale=0.5]{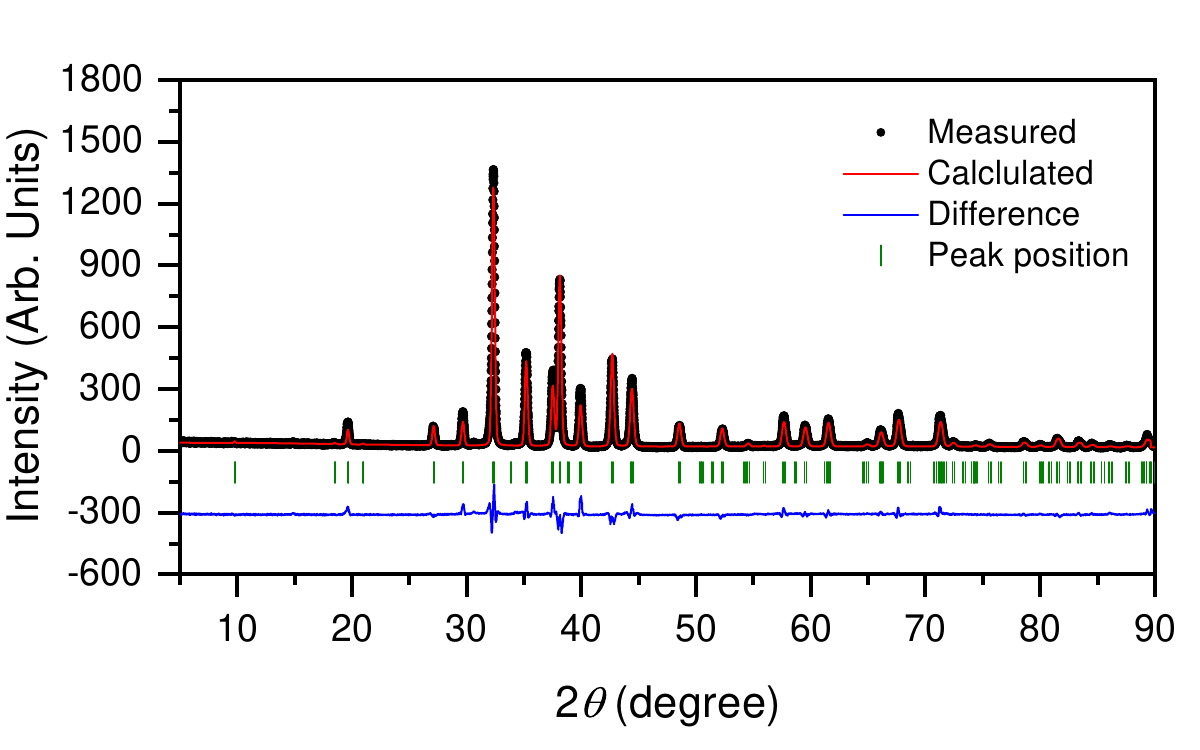}
    \caption{\textbf{Structural Characterization.} Rietveld refinement of the X-ray powder pattern of TmMn$_6$Sn$_{4.2}$Ga$_{1.8}$ measured at room temperature.} 
    \label{fig:FigS1}
    \end{center}
\end{figure}

\begin{table}[H]
\caption{Selected data from Rietveld refinement of powder
X-ray diffraction collected on ground crystals of TmMn$_6$Sn$_{4.2}$Ga$_{1.8}$.}\label{T1S}
\centering
\begin{tabular}{@{\hspace{.9cm}}l@{\hspace{.9cm}}l}	
 \hline
Space group & {\it P6/mmm} \\
 Unit cell $a$, $c$ (\AA) & 5.40848(8),~8.82625(17)\\
 \textit{R}$_{WP}$                     &     11.7 \%                               \\
  \textit{R}$_{B}$                     &     5.39 \%                               \\
   \textit{R}$_{F}$                           &     5.28  \%                               \\
   \hline
\end{tabular}

\begin{tabular}{c@{\hspace{0.3cm}}c@{\hspace{0.3cm}}c@{\hspace{0.3cm}}c@{\hspace{0.3cm}}c@{\hspace{0.3cm}}c}								
        Atom           & Position      & $x$         &	  $y$	          &	      $z$	    	        \\
		\hline
 Tm & 1$a$ & 0 & 0 & 0   \\														
 Mn & 6$i$ & 0 & 1/2 & 0.24080(19)   \\ 
 
Sn$_1$ & 2$e$ & 0 & 0 & 0.33495(18)   \\

Sn$_2$ & 2$d$ & 1/3 & 2/3 & 1/2  \\

Sn$_3$ & 2$c$ & 1/3 & 2/3 & 0  \\

Ga & 2$c$ & 1/3 & 2/3 & 0  \\

  \hline      					
\end{tabular}

\label{T1}
\end{table}

The crystal structure of TmMn$_6$Sn$_{4.2}$Ga$_{1.8}$ was confirmed through Rietveld refinement of powder X-ray diffraction (PXRD) data collected at room temperature using a Rigaku MiniFlex diffractometer. A representative single crystal was cleaned to remove residual surface flux and then ground into a fine powder for diffraction measurements. Rietveld refinements were performed using the FULLPROF software package \cite{Rodriguez-carvajal1993}. The refinement is shown in Fig. \ref{fig:FigS1}, with detailed crystallographic parameters summarized in Table \ref{T1}.

The analysis revealed that Ga substitution occurs specifically at the Sn$3$ site (Wyckoff position 2c), as modeling Ga occupancy at alternative Sn positions led to unphysical occupancies or poor refinement fits. This targeted substitution was further supported by chemical composition analysis using energy-dispersive X-ray spectroscopy (EDS), which confirmed a stoichiometry close to TmMn$_6$Sn$_{4.2}$Ga$_{1.8}$ with less than 7\% error. 

\section{Helicity Switching Analysis}\label{sec:1}

\begin{figure}[!ht]
\begin{center}
\includegraphics[scale=0.16]{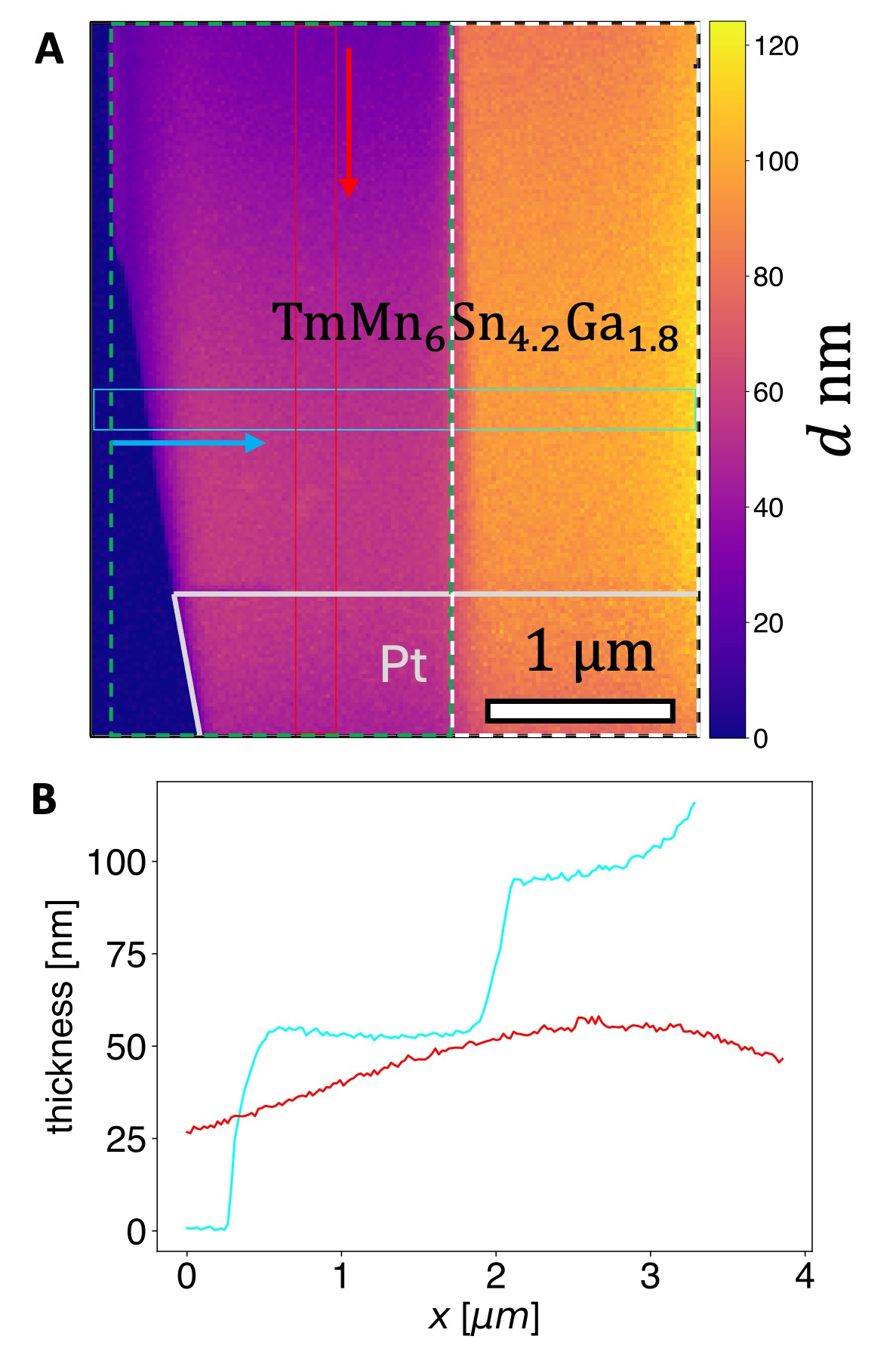}
    \caption{\textbf{Thickness mapping of a TmMn$_\textbf{6}$Sn$_{\textbf{4.2}}$Ga$_{\textbf{1.8}}$ lamella using electron energy loss spectroscopy (EELS), highlighting spatial variation across the sample}. (\textbf{A}) EELS image of the TEM sample studied, where (\textbf{B}) the plot shows vertical (red) and horizontal (blue) thickness profiles relative to distance from the lamella edge. The blue line displays a step-like variation due to the wedge-shaped preparation of the sample, while the red line remains relatively constant, representing the vertical thickness across the lamella. The average thicknesses in the thin and thick regions(excluding pixels with $d<5\ \textrm{nm}$) outlined by a green and white dashed rectangle, respectively, are  $\overline{d}_{\textrm{thin}}=43 \pm 10\ \textrm{nm}$ and $\overline{d}_{\textrm{thick}}=91 \pm 8\ \textrm{nm}$.} 
    \label{fig:FigS2}
    \end{center}
\end{figure}

We observed helicity switching in Type-I skyrmionic bubbles, which was found to be thickness-dependent. Using in situ temperature control in the LTEM, the switching behavior was tracked across a range of temperatures below the spin reorientation temperature (243 K). We confirmed helicity switching by observing the gradual, reversible rotation of magnetic domain contrast at zero magnetic field. The helicity-switching rate was inversely proportional to lamella thickness, suggesting a potential sensitivity to sample dimensions, likely due to variations in local anisotropy and dipolar interactions. Further analysis of this behavior is shown in Fig. \ref{Fig4}.

\section{Phase Diagram Details}\label{sec:2}

\begin{figure}[!ht]
\begin{center}
\includegraphics[scale=0.062]{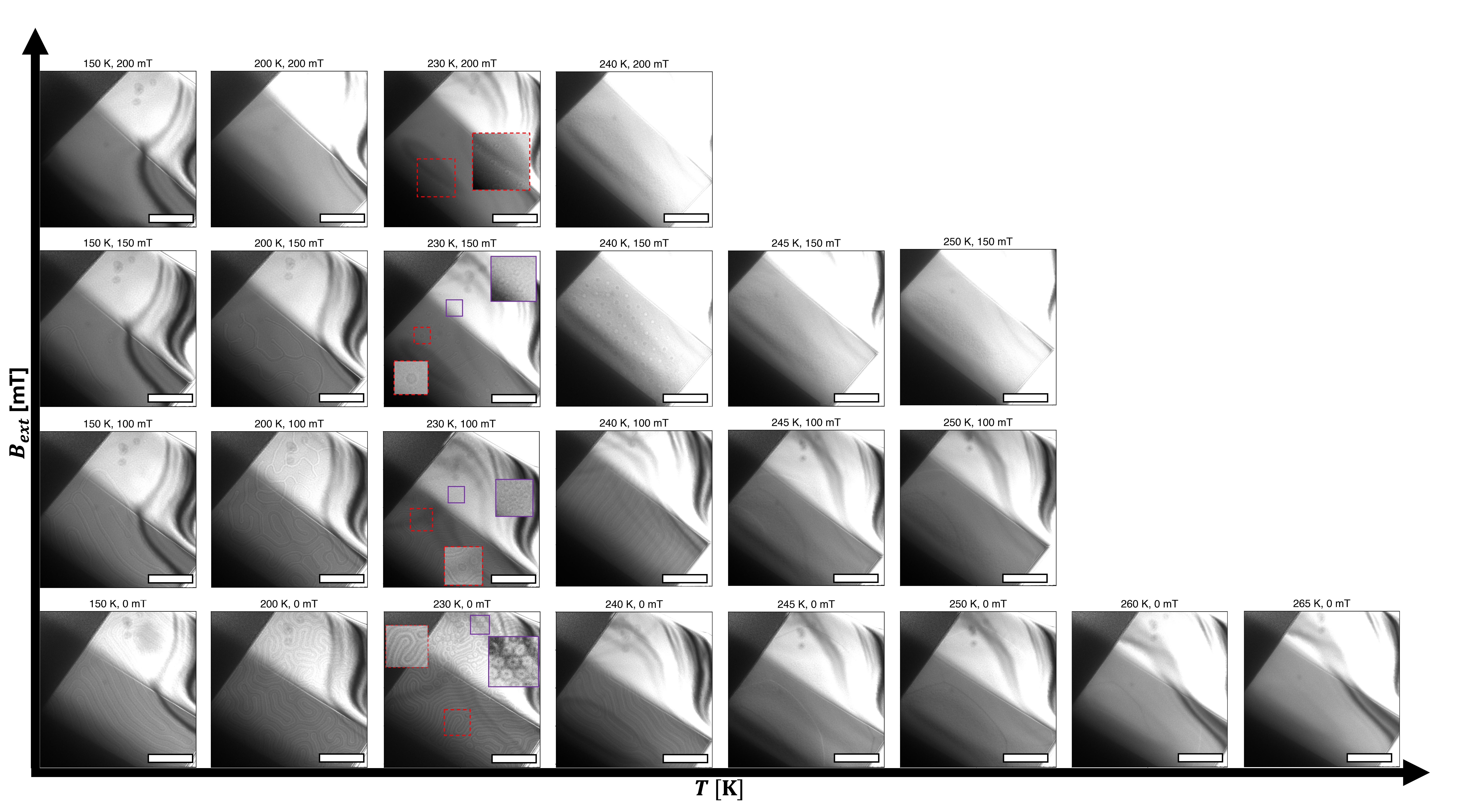}
    \caption{\textbf{Selected LTEM micrographs tabulated to illustrate the magnetic state diagram}. Images placed farther up the vertical axis correspond to elevated applied magnetic fields $B_{\textrm{ext}}$ whereas the horizontal axis corresponds to increasing temperature. The micrographs chosen span a temperature range of $T=150$ K to $T=265$ K and magnetic field range of $B_{\textrm{ext}}=0$ mT to $B_{\textrm{ext}}=200$ mT. Scale bars are $1\ \textrm{$\mu$m}$.}
    \label{fig:FigS3}
    \end{center}
\end{figure}

To construct the phase diagrams presented in the main text, magnetic field sweeps were conducted at $265 ~\textrm{K}$, $260 ~\textrm{K}$, $250 ~\textrm{K}$, $245 ~\textrm{K}$, $240 ~\textrm{K}$, ~$230\ \textrm{K}$, $200 ~\textrm{K}$, and $150 ~\textrm{K}$ from $0 ~\textrm{mT}$ to saturation. The diagrams were created based on observations of the real space magnetic contrast under increasing magnetic field along the c-axis, using LTEM imaging.  Fig. \ref{fig:FigS3} shows selected real space magnetic images acquired over the temperature and magnetic field ranges noted above. A clear change in magnetic contrast in the LTEM images in Fig. \ref{fig:FigS3} can be noted near the spin reorientation temperature, as described in the main text.

\begin{figure}[!ht]
\begin{center}
\includegraphics[scale=0.12]{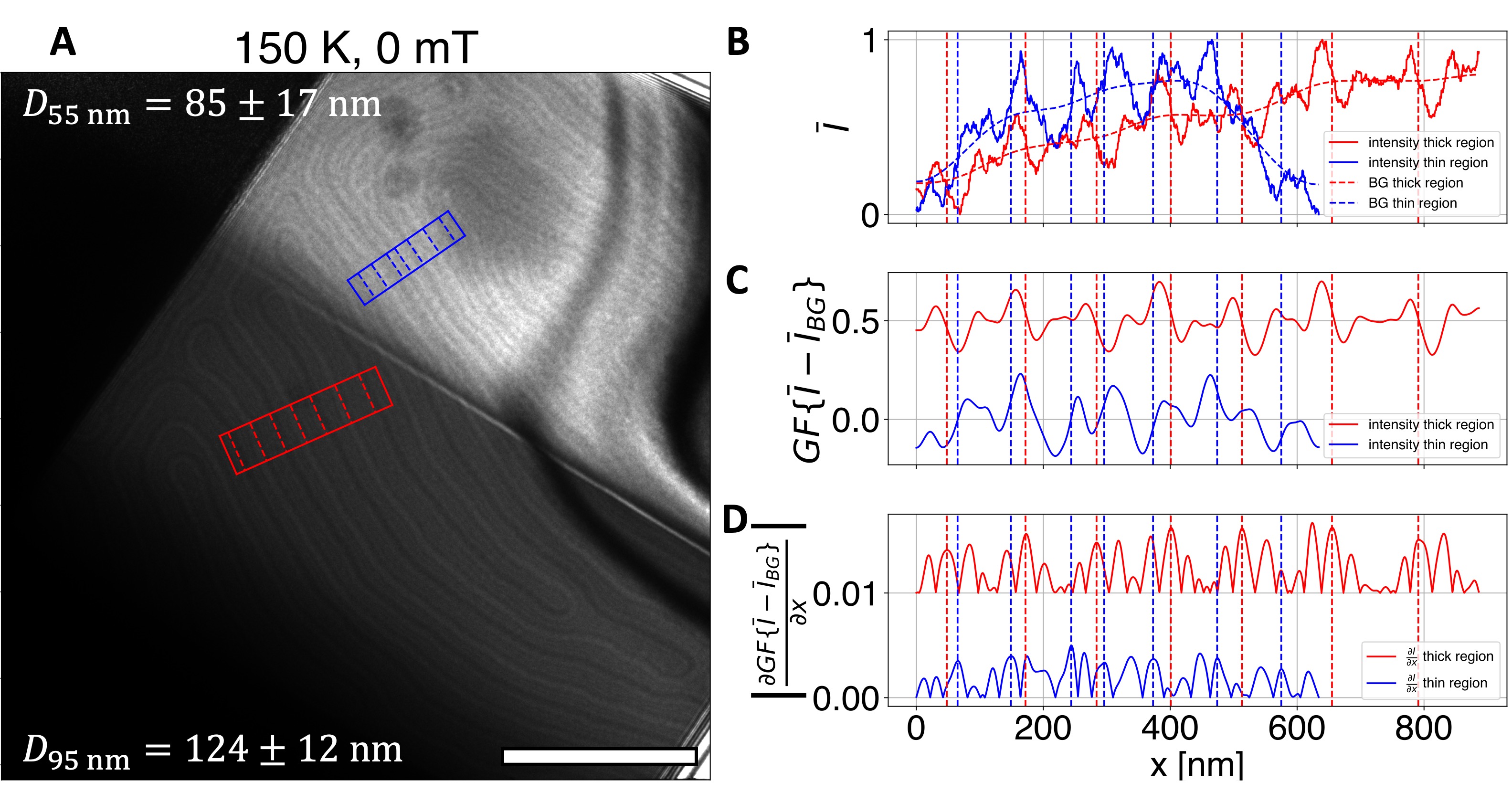}
    \caption{\textbf{Measurement and comparison of characteristic magnetic domain sizes in different thickness regions at $\textbf{T=150}\ \textrm{K}$ and zero field}. (\textbf{A}) LTEM micrograph of the TmMn$_6$Sn$_{4.2}$Ga$_{1.8}$ sample with stripey $180 ^\circ$ magnetic domains present in both the thin (lighter) and thick areas of the lamella. Scale bar is $1 \mu\textrm{m}$. (\textbf{B-D}) The (\textbf{B}) average intensity $\overline{I}$ of 50 adjacent one-dimensional slices from the thick and thin regions of the sample indicated  in (\textbf{A}) by red and blue rectangles, respectively. The low-frequency background intensity $\overline{I}_{\textrm{BG}}$, which we extract using a gaussian filter with $\sigma_{\textrm{GF}}=50\ \textrm{nm}$ is plotted by dashed lines. (\textbf{C}) Plot of the gaussian-filtered, background-subtracted average intensity $GF\{\overline{I} - \overline{I}_{\textrm{BG}}\}$ with $\sigma_{\overline{I}} = 7\ \textrm{nm}$. (\textbf{D}) The absolute valued derivative of $GF\{\overline{I} - \overline{I}_{\textrm{BG}}\}$ with respect to $x$. The seven domain wall positions are calculated as the $x$ location with the largest valued derivative within a $40 \textrm{nm}$ window of domain wall locations predetermined by eye. These domain wall $x$ positions for the thick and thin regions are marked with red and blue dashed vertical lines, respectively in (\textbf{B-D}).}
    \label{fig:FigS4}
    \end{center}
\end{figure} 

We measured the magnetic bubble densities by counting the number of bubbles in the thin and thick region of the lamella and dividing by the area of each region, $7.1\  \mu \textrm{m}^2$ and $6.8\  \mu \textrm{m}^2$, respectively at each magnetic field and temperature condition. As described in the main text, we then scaled the resulting density in the thicker region by 2.12: $\rho_{\text{bubbles}}^{*}(d=91\ \textrm{nm}) = D(d=91\ \textrm{nm}) \times D(d=43\ \textrm{nm}) / \rho_{\text{bubbles}}(d=91\ \textrm{nm}) \propto 91\ \textrm{nm} / 43\ \textrm{nm} = \rho_{\text{bubbles}}(d=91\ \textrm{nm}) \times 2.12$ to account for the difference in bubble density attributable the difference in thickness \cite{kittel1946theory}. In that work, the total energy of a thin film magnet with uniaxial anisotropy is derived to be $F = \sigma_{w} \left[ 2 \sqrt{2} + (d - D) / D \right] + \varepsilon_{\textrm{a}} D / 2$, where $\sigma_{w}$ is the surface energy density of the Bloch wall, $\varepsilon_{\textrm{a}}$ is the anisotropy energy density, $d$ is the film thickness, and $D$ is the characteristic domain size. In our case, Fig. \ref{fig:FigS4} provides a comparison of the domain sizes between the 43 nm and 91nm thick lamellae. Using the ground magnetic states after zero field cooling to $T=150\ \textrm{K}$, we determined that the domain sizes are $D_{43\ \textrm{nm}} = 85 \pm 17\ \textrm{nm}$ and $D_{91\ \textrm{nm}} = 124 \pm 12\ \textrm{nm}$, respectively. This results in a scaling factor of $2.12$, which matches our scaling value calculated using the EELS thickness measurements.

\section{Magnetic bubble reverification}

\begin{figure}[!ht]
\begin{center}
\includegraphics[scale=0.4]{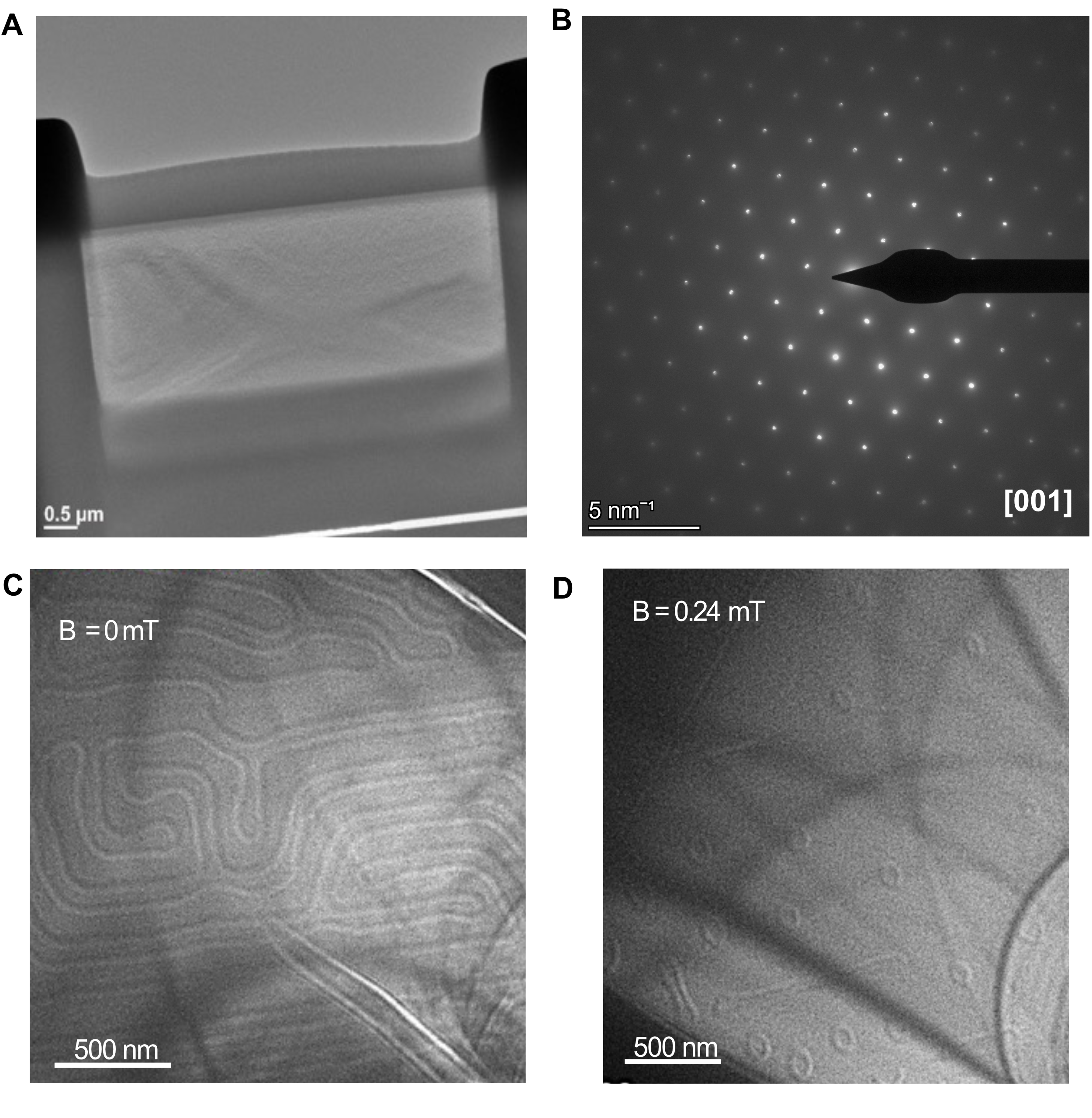}
    \caption{\textbf{Reproducibility of magnetic bubbles in a second growth batch of TmMn$_\textbf{6}$Sn$_{\textbf{4.2}}$Ga$_{\textbf{1.8}}$.} (\textbf{A}) SEM image of a FIB-prepared lamella oriented along the [001] direction, fabricated from a second independently grown TmMn$_6$Sn$_{4.2}$Ga$_{1.8}$ single crystal. (\textbf{B}) Selected area electron diffraction (SAED) pattern of the lamella, confirming orientation along the crystallographic $c$-axis. (\textbf{C}) Lorentz TEM image at 220 K and zero applied field reveals spontaneous stripe domains. (\textbf{D}) Magnetic bubbles emerge under an applied magnetic field of 0.24 mT.} 
    \label{fig:FigS6}
    \end{center}
\end{figure}

Magnetic bubble formation was re-verified on a second set of crystals from a new growth batch than those used for the LTEM and DPC-STEM presented in the main manuscript. A lamella of TmMn$_6$Sn$_{4.2}$Ga$_{1.8}$ (displayed in Fig. \ref{fig:FigS6}(A)) is prepared from bulk single crystals using a focused ion beam (FIB) system (Helios DualBeam, Thermo Fisher Scientific) with Ga$^+$ ions. The lamella was oriented along the crystallographic [001] axis, as confirmed by selected area electron diffraction (SAED), shown in Fig. \ref{fig:FigS6}(B). Final thinning and surface cleaning were performed at reduced ion energies of 5 kV and 2 kV to minimize surface damage and amorphization caused by ion milling.\\

Magnetic imaging was carried out using transmission electron microscopy in Lorentz mode (Spectra 300, Thermo Fisher Scientific)  at the Notre Dame Integrated Imaging Facility (NDIIF). A double-tilt liquid nitrogen cooling holder (Gatan 636) was employed to enable low-temperature studies of magnetic domain structures. LTEM images were acquired across a range of temperatures, with magnetic field applied along the [001] crystallographic direction. Magnetic field strength was modulated from 0 T to 0.3 T by finely tuning the objective lens current. At 220 K, with 0 applied magnetic field, stripe domains (Fig. \ref{fig:FigS6}(C)) emerged spontaneously. Upon increasing the magnetic field, these stripe domains progressively fragmented into the bubbles (Fig. \ref{fig:FigS6}(D)).

\end{document}